\documentclass[a4paper,11pt]{article}

\usepackage{jcappub} 

\usepackage[T1]{fontenc} 

\usepackage{amssymb,amsmath,amsfonts,amsbsy,graphicx,microtype,rotating,ulem}

\usepackage{color}

\def\be{\begin{equation}}
\def\ee{\end{equation}}
\def\ba{\begin{eqnarray}}
\def\ea{\end{eqnarray}}
\def\nn{\nonumber}

\title{\boldmath Scale-dependent CMB power asymmetry from primordial speed of sound and a generalized $\delta N$ formalism}

\author[a]{Dong-Gang Wang,}
\author[a,b,*]{Yi-Fu Cai,\note{Corresponding author.}}
\author[a]{Wen Zhao,}
\author[a]{Yang Zhang}

\affiliation[a]{CAS Key Laboratory for Researches in Galaxies and Cosmology, Department of Astronomy,  University of Science and Technology of China, Chinese Academy of Sciences, Hefei, Anhui 230026, China}
\affiliation[b]{Department of Physics, McGill University, Montr\'eal, Quebec H3A 2T8, Canada}

\emailAdd{wdgang@mail.ustc.edu.cn}
\emailAdd{yifucai@ustc.edu.cn}
\emailAdd{wzhao7@ustc.edu.cn}
\emailAdd{yzh@ustc.edu.cn}

\abstract{We explore a plausible mechanism that the hemispherical power asymmetry in the CMB is produced by the spatial variation of the primordial sound speed parameter. We suggest that in a generalized approach of the $\delta N$ formalism the local e-folding number may depend on some other primordial parameters besides the initial values of inflaton. Here the $\delta N$ formalism is extended by considering the effects of a spatially varying sound speed parameter caused by a super-Hubble perturbation of a light field. Using this generalized $\delta N$ formalism, we systematically calculate the asymmetric primordial spectrum in the model of multi-speed inflation by taking into account the constraints of primordial non-Gaussianities. We further discuss specific model constraints, and the corresponding asymmetry amplitudes are found to be scale-dependent, which can accommodate current observations of the power asymmetry at different length scales.}


\arxivnumber{1509.02541}

\begin{document}
\maketitle
\flushbottom
\section{Introduction}

The hemispherical power asymmetry in the cosmic microwave background (CMB) was indicated by the WMAP experiment years ago \cite{Eriksen:2003db, Hansen:2004vq, Eriksen:2007pc, Hoftuft:2009rq}. Later the Planck collaboration also reported this feature in the CMB fluctuations \cite{Ade:2013nlj, Ade:2015hxq} and provided an independent measurement of this anomaly. Phenomenologically, such a power asymmetry can be modeled as a dipolar modulation of a statistically isotropic CMB sky in terms of temperature fluctuations in a specific direction of the unit vector $\hat{p}$. Translated into the expression of the primordial power spectrum, the modulation required to explain this asymmetry can be written as a spatially-varying power spectrum,
\begin{eqnarray}\label{P_kr}
 P_{\zeta}(k, \vec{r}) = P_{\zeta}(k) \left[ 1 + 2A(k) ~\hat{p} \cdot ({{\bf x}-{\bf x}_0}) / {x_{ls}} \right]~,
\end{eqnarray}
where $P_{\zeta}(k)$ is the isotropic spectrum, $A(k)$ characterizes the amplitude of dipolar asymmetry, $x_{ls}$ is the comoving distance to the last scattering surface and ${\bf x}_0$ represents our current location.

The amplitude of the dipolar asymmetry is given by $A=0.072 \pm 0.022$  for the CMB power with $\ell \lesssim 60$  \cite{Ade:2013nlj}. However, the asymmetry does not necessarily exist at smaller length scales. Particularly, the constraint from the Sloan Digital Sky Survey quasar sample \cite{Hirata:2009ar} requires $A<0.0153$ ($99\%$ C.L.) for the power asymmetry oriented in the direction of the CMB dipole in which the typical wavenumber is $k \sim 1 {\rm Mpc}^{-1}$.
And analyzing CMB data only also indicates a small asymmetry amplitude at large $\ell$ \cite{Flender:2013jja, Aiola:2015rqa}.  Thus, any model that accounts for the CMB power asymmetry has to produce a strong scale dependence so that it can be in agreement with the above observational constraints. In order to explain the observed scale-dependent CMB anisotropy various theoretical models have been proposed in the literature \cite{Land:2005cg, Boehmer:2007ut, Campanelli:2007qn, Erickcek:2008sm, Erickcek:2008jp, Kahniashvili:2008sh, Carroll:2008br, Watanabe:2010fh, Chang:2013vla, Wang:2013lda, McDonald:2013aca, Mazumdar:2013yta, Cai:2013gma, Liu:2013kea, Liu:2013iha, Kanno:2013ohv, McDonald:2014lea, Kothari:2015tqa, Kohri:2013kqa, Liddle:2013czu, Dai:2013kfa, Lyth:2013vha, Namjoo:2013fka, Abolhasani:2013vaa, Firouzjahi:2014mwa, Lyth:2014mga, Namjoo:2014nra, Kobayashi:2015qma, Jazayeri:2014nya, Mukherjee:2015mma, Mukherjee:2015wra, Byrnes:2015dub}. 

As pointed out in \cite{Erickcek:2008sm}, a single-field slow-roll inflation model cannot generate such an asymmetry without violating the constraints of homogeneity in the universe. The same paper also proposed a so-called Erickcek-Kamionkowski-Carroll (EKC) mechanism based on a curvaton model \cite{Lyth:2001nq, Mollerach:1989hu, Linde:1996gt, Enqvist:2001zp} and thus can explain this anomaly without violating the homogeneity constraint. However, the original model is inconsistent with the quasar bound since the signature is scale-independent. Also, the model leads to a large value of non-Gaussianity which has been ruled out by Planck \cite{Ade:2013ydc}. Instead, a improved curvaton model in which the curvaton decay takes place after dark matter freezes out was studied \cite{Erickcek:2009at}.

When dealing with cosmological perturbations in a system involving multiple scalar fields, it turns out that the $\delta{N}$ formalism \cite{Starobinsky:1986fxa, Salopek:1990jq, Sasaki:1995aw, Lyth:2004gb, Lyth:2005fi} is very powerful, particularly when describing nonlinear terms. The application of the $\delta{N}$ formalism in the study of the CMB power asymmetry has been carried out in the literature \cite{Erickcek:2008sm, Lyth:2013vha, Namjoo:2013fka, Abolhasani:2013vaa, Kohri:2013kqa, Kanno:2013ohv, Lyth:2014mga, Namjoo:2014nra, Kobayashi:2015qma}.
It was found in these studies that there may exist certain discrepancy if the nonlinear contributions to the quadrupole were not treated in a full analysis, as was pointed out in \cite{Kobayashi:2015qma}.

In spite of all those models, we suggested that the hemispherical power asymmetry may be generated by a modulation of primordial speed of sound parameter during inflation as studied in Ref.~\cite{Cai:2013gma}. In the inflation model of K-essence field \cite{ArmendarizPicon:1999rj}, the propagation of the field fluctuation was found to be governed by a speed of sound parameter $c_s$. This parameter usually equals to unity when the kinetic term is of standard form. However, when the kinetic term is nontrivial, the sound speed parameter could be smaller than unity and yields a different squeezing process for the field fluctuation and thus modifies the primordial power spectrum of the curvature perturbation. Further, the model of {\it multi-speed inflation} is constructed by a collection of scalar fields with such nonstandard kinetic terms \cite{Cai:2009hw, Cai:2008if}. Then, in this model there is one corresponding sound speed for each field, and therefore the fluctuations of these fields do not propagate synchronously.

In the present work we extend the $\delta{N}$ calculation by considering the spatially varying primordial sound speed parameter of the inflaton field. Such a modulation of the sound speed can be obtained by the super-Hubble perturbation of a second light field, which is dubbed the EKC effect. This mechanism is implemented in the model of the multi-speed inflation. In Section II, we give a brief review of this model and the $\delta{N}$ formalism. In Section III we illustrate the feasibility of our mechanism by giving a detailed calculation of the asymmetry amplitude $A(k)$. Afterwards, in Section IV we consider a specific model and numerically examine the effect of the modulation of sound speed parameter. Finally a scale-dependent asymmetry amplitude is obtained to confront the observational data. Section V includes the conclusion with a discussion.

\section{A brief review of multi-speed inflation and $\delta N$ formalism}

Our starting point is based on the multi-speed inflation model, which was proposed in Refs.~\cite{Cai:2009hw, Cai:2008if}. In a simple version of the model we consider that the universe is driven by two scalar fields minimally coupled to Einstein gravity at early times, with one being a canonical field $\chi$ and the second being a K-essence field $\phi$ which has a non-canonical kinetic term $P(X,\phi)$ with $X = -\frac{1}{2} g^{\mu\nu} \partial_\mu\phi \partial_\nu\phi$. This K-essence type of the scalar field may arise from a low energy effective field description of string theory or other more fundamental theories \cite{ArmendarizPicon:1999rj}. For instance, the non-canonical kinetic term could be of the Dirac-Born-Infeld-like (DBI) form, and accordingly, the Lagrangian density can be written as
\be \label{lagrangian}
\mathcal{L}=P(X,\phi,\chi)
-\frac{1}{2}\partial_\mu\chi\partial^\mu\chi-V(\phi, \chi) ~,
\ee
with
\be
P(X,\phi,\chi) = \frac{1}{f(\phi,\chi)} \Big[ 1- \sqrt{1+f(\phi,\chi) \partial_\mu\phi \partial^\mu\phi} \Big] ~.
\ee
The coefficient $f(\phi,\chi)$ inside the square root function is a warping factor for the scalar field $\phi$.

Here we assume that $\phi$ plays the role of inflaton while $\chi$ makes no contribution to the background dynamics\footnote{For instance, this model was applied in Refs.~\cite{Cai:2010, Cai:2009} in order to realize a viable curvaton mechanism and was investigated in \cite{Emery:2013yua} to produce negative-valued local non-Gaussianities.}. In this model a distinct property is that the speed of sound parameters of those two fields can behave differently. For $\chi$, the sound speed is simply unity since its kinetic term is canonical. For $\phi$, however, the square of the sound speed is given by
\be \label{cs}
c_s^2\equiv\frac{P_{,X}}{P_{,X}+2XP_{,XX}}=1-2f(\phi,\chi)X ~,
\ee
where $P_{,X}$ denotes the partial derivative of $P$ with respect to $X$. Note that $c_s$ is time-dependent and affected by the variation of scalar field $\chi$. In this model the sound speed parameter is expected to be related to the primordial non-Gaussianities via the following expression \cite{Chen:2006nt}
\be \label{NG}
f_{NL}^{DBI} =-\frac{35}{108} \Big( \frac{1}{c_s^2}-1 \Big) ~.
\ee
Then one can observe that a small value of the sound speed parameter can lead to a large amount of primordial non-Gaussianity, while $c_s\sim1$ corresponds to $f_{NL}\ll1$. Hence, the model can be well constrained by cosmological observations such as the Planck data \cite{Ade:2015ava, Ade:2015lrj}.

\subsection{Background dynamics}

To study the background dynamics, we vary the Lagrangian with respect to the metric and then obtain the energy-momentum tensor as follows,
\be \label{emt}
T^{\mu\nu} = g^{\mu\nu}(P-V) +P_{,X} \nabla^\mu\phi \nabla^\nu\phi ~.
\ee
Moreover, varying the Lagrangian with respect to the inflaton yields the following generalized Klein-Gordon equations,
\begin{eqnarray}
 \nabla_\mu({P}_{,X} \nabla^\mu\phi) +{P}_{,\phi}-V_{,\phi}=0 ~,
\end{eqnarray}
where ${P}_{,\phi}$ denotes the partial derivative of $P$ with respect to the scalar $\phi$.

Considering a spatially flat Friedmann-Robertson-Walker (FRW) spacetime with its metric
\begin{eqnarray}
ds^2=-dt^2+a^2(t)d{\bf x}^2 ~,
\end{eqnarray}
we can read the energy density and pressure of the above cosmological system from the energy-momentum stress tensor
\be \label{rho}
\rho=\frac{\dot\phi^2}{c_s}-P+V,\ \ \ \ \ \ \ \ p=P-V ~.
\ee
The equations of motion for the scalar fields now can be expressed as
\be \label{eom}
\ddot \phi+3H\dot\phi-\frac{\dot c_s}{c_s}\dot\phi-c_s\frac{\partial P}{\partial \phi}+c_s\frac{\partial V}{\partial \phi}=0 ~,
\ee
where the Hubble expanding rate $H\equiv\dot a/a$ has been introduced. The second Friedman Equation is given by
\be \label{Fri}
\dot H = -\frac{1}{2 M_{pl}^2}(\rho+p) = -\frac{\dot\phi^2}{2 M_{pl}^2c_s} ~,
\ee
where $M_{pl}$ is the reduced Planck mass, which is defined as $1/\sqrt{8\pi G}$. Moreover, we can introduce the following three slow roll parameters
\be \label{slowroll}
\epsilon\equiv-\frac{\dot H}{H^2} ~,~~ \eta\equiv\frac{\dot\epsilon}{H\epsilon} ~,~~ s\equiv\frac{\dot c_s}{H c_s} ~,
\ee
and they respectively characterize the variation of $H$, $\epsilon$ and $c_s$ in one Hubble time. To ensure a sufficiently long and stable inflationary phase, these parameters are demanded to be much smaller than $1$ during inflation. In addition, combining Eqs. (\ref{Fri}) and (\ref{slowroll}), we get the following relation
\be \label{hphi}
\frac{H^2}{\dot\phi^2} = \frac{1}{2 M_{pl}^2\epsilon c_s} ~,
\ee
which is very useful in the analysis of the generalized $\delta N$ formalism in next section.

\subsection{$\delta N$ formalism and the isotropic spectrum}

Having the background dynamics of multi-speed inflation, we review the basic idea of the $\delta N$ formalism in this subsection. Afterwards, we use this approach to calculate the power spectrum of primordial curvature perturbation $P_{\zeta}(k)$. During inflation the regions of super-Hubble scales may be regarded as many ``separate universes''. Each of these regions is approximately homogeneous and isotropic, and thus, the corresponding evolution can be locally described by an FRW one. However, due to quantum fluctuations, the primordial parameters, such as the initial value of inflaton, are different in these regions, which can lead to different expansion behavior. We may define $N({\bf x},t)$ as the local e-folding number from a spatially flat slice to a uniform-density slice at time $t$. Then the curvature perturbation of this region can be interpreted as the difference between $N({\bf x},t)$ and the background e-folding number, i.e.
\be
\zeta({\bf x},t)=N({\bf x},t)-N(t)=\delta N({\bf x},t) ~,
\ee
which is known as the $\delta N$ formalism and is formulated as follows,
\be \label{deltaN}
\delta N=\frac{\partial N}{\partial \phi_i}\delta\phi_i+\frac{1}{2}\frac{\partial^2 N}{\partial \phi_i \partial \phi_j}\delta\phi_i\delta\phi_j +... ~,
\ee
where a subscript $_i$ denotes the evaluation along the $i$-th unperturbed field trajectory.

Here $\delta\phi_i$ corresponds to the $i$-th field fluctuations whose wave lengths are smaller than the current observable universe, i.e. $k>x_{ls}^{-1}$. Since in our model the background evolution is independent of $\chi$, there is $\frac{\partial N}{\partial \chi}=0$. In this case, the relation between $N$ and $\phi$ is simply determined by the background dynamics of inflation
\be \label{Nphi}
\frac{\partial N}{\partial \phi}=-\frac{H}{\dot\phi} ~.
\ee
To leading order in the expansion of Eq. (\ref{deltaN}), the power spectrum of curvature perturbation turns to be
\be
P_\zeta(k)=\frac{k^3}{2\pi^2}\left|\frac{\partial N}{\partial \phi}\right|^2|\delta\phi_k|^2 ~.
\ee
As is well known, $P_{\delta\phi}\equiv \frac{k^3}{2\pi^2}|\delta\phi_k|^2\simeq\left(\frac{H}{2\pi}\right)^2$ in a quasi-de Sitter spacetime. Then using Eqs. (\ref{Nphi}) and (\ref{hphi}), we derive
\be \label{Nphi2}
\left|\frac{\partial N}{\partial \phi}\right|^2=\frac{1}{2 M_{pl}^2\epsilon c_s} ~,
\ee
and thus, the isotropic power spectrum in our model is expressed as,
\be
P_\zeta=\frac{H^2 }{8\pi^2 M_{pl}^2\epsilon c_s} ~.
\ee

\section{Asymmetric spectrum caused by the spatial varying sound speed}

In this section we firstly extend the $\delta N$ formalism by taking into account the effects of a varying speed of sound parameter. Then we use this generalized approach to analyze the generation of power asymmetry caused by the spatially varying sound speed.

\subsection{A generalized $\delta N$ formalism}

As was reviewed in the previous section, the regular $\delta N$ formalism considers only the direct dependence of the e-folding number $N$ on the inflaton $\phi$. However, there exist some primordial parameters that may vary with the background evolution during inflation. In our case, from Eqs. (\ref{eom}) and (\ref{Fri}), one can observe that the primordial sound speed parameter $c_s$ is such a parameter. Thus we suggest that besides the initial value of inflaton $\phi({\bf x})$, $c_s$ also influences the local e-folding number through a generalized formalism: $N({\bf x},t)=N(\phi({\bf x}),c_s({\bf x}),t)$. Accordingly, the $\delta N$ formula in Eq. (\ref{deltaN}) can be extended as follows,
\begin{align} \label{deltaN2}
\zeta = \frac{\partial N}{\partial \phi}\delta\phi +\frac{\partial N}{\partial c_s}\Delta c_s +\frac{1}{2}\frac{\partial^2 N}{\partial \phi^2} \delta\phi^2  +\frac{\partial^2 N}{\partial c_s \partial \phi} \delta\phi\Delta c_s +\frac{1}{2}\frac{\partial^2 N}{\partial c_s^2} \Delta c_s^2 + ...
\end{align}
up to second order. As before, the derivatives of $N$ with respect to $c_s$ are evaluated at the value of the background sound speed $c_s^b$. In the following, we use the extended $\delta N$ formalism to analyze the effects of a spatially varying sound speed parameter and apply this approach to calculate the corresponding asymmetric primordial power spectrum.

\subsection{The spatial variation of the sound speed}

Let us consider the situation that the sound speed $c_s$ is varying spatially as follows,
\be
c_s({\bf x},t) = \overline{c_s}(t) +\Delta c_s({\bf x},t) ~.
\ee
According to Eq. (\ref{cs}), this modulation of the sound speed can be achieved by considering the long wavelength field perturbation $\Delta \chi({\bf x},t)$ through
\be \label{partialcs}
\Delta c_s({\bf x},t)=\frac{\partial c_s}{\partial \chi}\Delta\chi({\bf x},t) ~.
\ee
From Eqs. (\ref{cs}) and (\ref{NG}), we have
\be
\frac{\partial c_s}{\partial \chi} = -\frac{f_{,\chi}}{2f} \Big( c_s-\frac{1}{c_s} \Big) ~.
\ee
For $f_{NL}\ll1$ (i.e. $c_s\sim1$), the leading order of Eq. (\ref{partialcs}) reduces to the one developed in Ref. \cite{Cai:2013gma}, which is given by
\[
c_s = \overline{c_s} \Big[ 1+\frac{54}{35}f_{NL} \frac{f_{,\chi}}{f}\Delta\chi \Big] ~.
\]
Different from the sub-Hubble perturbation $\delta\chi$, here $\Delta\chi({\bf x},t)$ is a super-Hubble mode with specific direction and wavelength. According to the Grishchuk-Zel'dovich (GZ) effect \cite{GZ}, the very large scale mode $\Delta \chi$ could modify the spectrum in the observable universe.

Following the treatment in Ref. \cite{Kobayashi:2015qma}, we assume that $\Delta \chi$ only exists at single wave number ${\bf k}_L$ and treat it as a non-stochastic quantity. Thus it can be expressed as
\be \label{dchi}
\Delta\chi({\bf x},t)=B(t)\cos({\bf k}_L\cdot{\bf x}) ~,
\ee
where ${ k}_L\ll x_{ls}^{-1}$. We further assume that we are located at ${\bf x_0}$, then there is
\be \label{expand}
{\bf k}_L\cdot{\bf x} ={\bf k}_L\cdot{\bf x_0} +{\bf k}_L\cdot({\bf x}-{\bf x_0}) ~.
\ee
Considering the limit of ${\bf k}_L\cdot({\bf x}-{\bf x_0})\ll1$, one can expand Eq. (\ref{dchi}) to be
\be
\Delta\chi({\bf x},t) = B(t) \big[ \cos\theta - \sin\theta ~{\bf k}_L\cdot({\bf x}-{\bf x_0}) \big] ~,
\ee
with $\theta\equiv{\bf k}_L\cdot{\bf x_0}$. As a result, the spatially varying part of the sound speed parameter takes the form of
\begin{align} \label{deltacs}
\Delta c_s({\bf x},t) = \overline{\Delta c_s}(t)+\widetilde{\Delta c_s}({\bf x},t)= \frac{\partial c_s}{\partial \chi} B(t) \big[ \cos\theta - \sin\theta ~{\bf k}_L\cdot({\bf x}-{\bf x_0}) \big] ~.
\end{align}

\subsection{The spatially varying spectrum}

Having the above consideration, now we turn to calculate the asymmetric spectrum.
In the Fourier space, the extended $\delta N$ formalism (\ref{deltaN2}) yields
\begin{align} \label{deltaNk}
 \zeta_{\bf k} =& \frac{\partial N}{\partial \phi}\delta\phi_{\bf k} +\frac{\partial N}{\partial c_s}\Delta c_s({\bf k}) +\frac{1}{2} \frac{\partial^2 N}{\partial \phi^2} \int\frac{d^3{\bf p}}{(2\pi)^3} \delta\phi_{\bf p} \delta\phi_{{\bf k-p}} +\frac{\partial^2 N}{\partial c_s \partial \phi} \int\frac{d^3{\bf p}}{(2\pi)^3} \delta\phi_{\bf p} \Delta c_s({\bf k-p}) \nn\\
 & +\frac{1}{2}\frac{\partial^2 N}{\partial c_s^2} \int\frac{d^3{\bf p}}{(2\pi)^3} \Delta c_s({\bf p}) \Delta c_s({\bf k-p}) ~,
\end{align}
at second order.

The small scale perturbations $\delta\phi_{\bf k}$ are approximated to be of highly Gaussian distribution, and thus, we have the following expressions,
\begin{align}
 \langle\delta\phi_{\bf k}\rangle = 0 ~, ~~
 \langle\delta\phi_{\bf k}\delta\phi_{\bf k'}\rangle = (2\pi)^3 \frac{2\pi^2}{k^3} \Big( \frac{H}{2\pi} \Big)^2 \delta({\bf k+k'}) ~,~~
 \langle\delta\phi_{\bf k}\delta\phi_{\bf k'}\delta\phi_{\bf k''}\rangle = 0 ~,
\end{align}
where we have defined $k\equiv|\bf k|$. As the large scale perturbation $\Delta \chi$ is not stochastic, it has no correlation
with $\delta\phi$ and itself. Using these relations and for $k_L\ll k$, one obtains the asymmetric power spectrum
\begin{align}
 P_{\zeta}(k,{\bf r}) = \frac{k^3}{2\pi^2} \frac{\langle \zeta_{{\bf k_1}} \zeta_{{\bf k_2}} \rangle}{(2\pi)^3} = \Big[ \frac{\partial N}{\partial \phi} +\frac{\partial^2 N}{\partial c_s \partial \phi} \Delta c_s({\bf x},t) \Big]^2 \Big( \frac{H}{2\pi} \Big)^2 ~.
\end{align}
To apply Eq. (\ref{expand}), we can expand the power spectrum to the leading order of ${\bf k}_L\cdot{\bf (x-x_0)}$ as follows,
\begin{align}
 P_{\zeta}(k,{\bf x}) = \Big[ \frac{\partial N}{\partial \phi} +\frac{\partial^2 N}{\partial c_s \partial \phi} \overline{\Delta c_s}(t) \Big]^2 \Big(\frac{H}{2\pi}\Big)^2\Big[ 1+\frac{2({\partial^2 N}/{\partial c_s \partial \phi})}{\frac{\partial N}{\partial \phi} +\frac{\partial^2 N}{\partial c_s \partial \phi} \overline{\Delta c_s}(t)}\widetilde{\Delta c_s}({\bf x},t) \Big] ~.
\end{align}
As indicated in Eq. (\ref{deltacs}), $\overline{\Delta c_s}$ is only an overall shift of the background value. Therefore, we can rewrite the background sound speed parameter during inflation to be $c_s^b=\overline{c_s}+\overline{\Delta c_s}$ and then derive
\be
\frac{\partial N}{\partial \phi} +\frac{\partial^2 N}{\partial c_s \partial \phi} \overline{\Delta c_s}(t) = \left. \frac{\partial N}{\partial \phi} \right|_{c_s^b} ~.
\ee
As a result, the asymmetric power spectrum can be further simplified as the same form of Eq. (\ref{P_kr}).

Moreover, the amplitude of the power asymmetry in this case is expressed as
\be \label{Ak}
A(k) = -(k_L x_{ls}) {\frac{\partial^2 N}{\partial c_s \partial \phi}} \Big( \frac{\partial N}{\partial \phi} \Big)^{-1} \frac{\partial c_s}{\partial \chi} B(t)\sin\theta ~.
\ee
Recall that, according to Eq. (\ref{Nphi}), the background dynamics of our model yields the following relation
\be \label{Nphi3}
\frac{\partial N}{\partial \phi}=-\frac{H}{\dot\phi}=-\frac{1}{ M_{pl}\sqrt{2\epsilon c_s}} ~.
\ee
Furthermore, by taking the partial derivative of the above formula with respect to $c_s$ and considering the variation of the slow roll parameter $\epsilon$, we can get another useful relation
\be \label{Nphics}
 \frac{\partial^2 N}{\partial c_s \partial \phi} = \frac{1}{2 M_{pl} \sqrt{2\epsilon }c_s^{3/2}} \Big( 1 +\frac{\eta}{s} \Big) ~,
\ee
where we have already applied the definition of slow roll parameters (\ref{slowroll}) and the following relation
\be
\frac{\partial \epsilon}{\partial c_s}=\frac{\dot\epsilon }{\dot c_s}=\frac{\epsilon \eta}{c_s s} ~.
\ee
Finally, substituting (\ref{Nphi3}) and (\ref{Nphics}) into (\ref{Ak}), we obtain the amplitude of the power asymmetry of the following form
\ba \label{Ak2}
 A(k) = k_L x_{ls} \Big( 1+\frac{\eta}{s} \Big) \frac{f_{,\chi}}{4f} \Big( 1-\frac{1}{c_s^2} \Big) B(t)\sin\theta ~.
\ea

Before moving to the next section, we would like to remark that in the EKC mechanism the asymmetry amplitude is related to the local non-Gaussianity via
\be
f^{local}_{NL} = \frac{5}{6}\frac{\partial^2N}{\partial\phi^2} \Big/ \Big( \frac{\partial N}{\partial\phi} \Big)^2 ~.
\ee
Thus the observational constraints on primordial non-Gaussianity may result in the unnatural fine tuning of model parameters or a running behavior of $f^{local}_{NL}$ if the power asymmetry is due to the inflaton field\cite{Kobayashi:2015qma, Byrnes:2015dub}. In our approach the asymmetry amplitude (\ref{Ak}) is not related to the primordial local non-Gaussianity, and thus, there is no necessity to require any fine tuning or a running $f^{local}_{NL}$.

\section{Specific example and confronting observations}

In this section, we use specific implementation of the multi-speed model to present how our mechanism provides a successful explanation for the hemispherical power asymmetry and use the observational data to constraint the model.

Notice that the constraint of Planck on the primordial non-Gaussianity of the DBI model only gives $f^{DBI}_{NL}=2.6\pm61.6$ and $c_s>0.069$ \cite{Ade:2015ava}, thus there is a large parameter range that we could explore in our case. For example, if we demand $c_s\simeq1$ (i.e. $f_{NL}\ll 1$ ) and $\eta/s\ll1$, Eq. (\ref{Ak2}) reduces to
\be \label{preA}
|A(k)|=\frac{54}{70}\frac{f_{,\chi}}{f}f_{NL}\frac{H}{2\pi}E,
\ee
with $\frac{H}{2\pi}E=k_L r_{ls}B(t)\sin\theta$. This result has been fully discussed in the previous work \cite{Cai:2013gma}.

Now let us take a more specific example based on Ref. \cite{Cai:2009hw} which indicates the case with $\eta\sim s$ and $c_s\ll1$. As we have pointed out, there are two undefined functions in the original Lagrangian, the warping factor $f(\phi, \chi)$ and the potential $V(\phi, \chi)$. If we consider an AdS warping throat, the warping factor can be expressed as
\be
f(\phi,\chi) = f(\phi) +f(\chi) =\frac{\lambda}{\phi^4}+\gamma \chi^p ~,
\ee
while the IR type potential takes the following form
\be
V(\phi) = V_0 - \frac{1}{2} m^2\phi^2 ~.
\ee
Under the relativistic limit of inflaton with small sound speed and the requirement $|Ht|\gg1$ during inflation, an approximate solution of Eq. (\ref{eom}) is given by \cite{Cai:2009hw, Cai:2008if}
\be
\phi = -\frac{\sqrt{\lambda}}{t} \Big( 1-\frac{9H^2}{2m^4t^2} +... \Big) ~,
\ee
where the beginning of inflation is set at $t \rightarrow -\infty$ and higher order terms are suppressed by $\frac{1}{Ht}$.
By Eq. (\ref{cs}) the leading order of the sound speed parameter takes the form of
\be \label{csevol}
c_s \simeq -\frac{3H}{m^2t} ~.
\ee

Afterwards, one can calculate the slow roll parameters at leading order, which are given by
\begin{align}
 \epsilon &= \frac{\dot\phi^2}{2c_sH^2M_{pl}^2} =\frac{\lambda m^2}{6M^2_{pl}H^3t^3} +\mathcal{O}(\frac{1}{H^4t^4}) ~, \nn\\
 \eta &\equiv \frac{\dot\epsilon}{H\epsilon} =-\frac{3}{Ht} +\mathcal{O}(\frac{1}{H^2t^2}) ~, \nn\\
 s &\equiv \frac{\dot c_s}{H c_s} =-\frac{1}{Ht} +\mathcal{O}(\frac{1}{H^2t^2}) ~.
\end{align}
This indicates that $\eta/s \approx 3$ in the relativistic limit. Thus the asymmetry amplitude in Eq. (\ref{Ak2}) yields
\be \label{Ak3}
|A(k)|\simeq
\frac{k_L x_{ls}}{\chi(t_k)}\frac{p}{c_s(t_k)^2}B(t)\sin\theta.
\ee
From Eq. \eqref{Ak3}, we can see that there exists a $k$-dependent behavior of the asymmetry amplitude which originates from the evolution of the $\chi$ field  via the sound speed  during inflation. The amplitude of the power spectrum is fixed when the mode $k$ exits the Hubble radius, i.e. $k<aH$. The perturbation modes of different scales correspond to different Hubble-exit time $t_k$, while the evolution of $\chi$  as well as $c_s$  gives different field values at $t_k$. Thus this can be used to provide a natural explanation of the observational fact that $A(k)$ is significant at cosmological scale but turns to be negligible at the Mpc scale.

For the amplitude of the super-horizon perturbation $\Delta\chi$, it is unclear to us whether this amplitude is enhanced or not. Since there is no observational evidence to imply any enhancement of perturbation modes at super-Hubble scales, it is reasonable to assume that $\Delta\chi$ has the same amplitude with $\delta\chi$ and $\delta\phi$.  Hence, we can approximately take $k_L x_{ls}B(t) \sim \frac{H}{2\pi} =\sqrt{2\epsilon c_s} P_\zeta^{1/2} M_{pl} \sim \sqrt{\frac{1-n_s}{2} c_s} P_\zeta^{1/2} M_{pl}$. Then Eq. (\ref{Ak3}) becomes
\be \label{Ak4}
|A(k)|\simeq\sqrt{\frac{1-n_s}{2}}\frac{P_\zeta^{1/2}p M_{pl}\sin\theta}{\chi(t_k)c_s(t_k)^{3/2}} ~.
\ee
According to the latest Planck data \cite{Ade:2015ava, Ade:2015lrj}, we learn that $n_s = 0.968 \pm0.006$, $c_s \geq 0.069$ and $\ln(10^{10}P_\zeta) = 3.062 \pm 0.029$ at the $68\%$ confidence level.
In the following, we set $c_s(t_s)=0.1$ at the Planck pivot scale $k_s=0.05{\rm Mpc}^{-1}$. Then by applying Eq. (\ref{NG}) we immediately get $f_{NL}^{DBI}=-32$, which is consistent with the constraint of the Planck data, which is $f^{DBI}_{NL}=2.6\pm61.6$ \cite{Ade:2015ava}. Making use of Eq. (\ref{csevol}) and horizon exit condition $a(t_k)H(t_k)=k$, we get the $k$-dependent sound speed, which is expressed as
\be
c_s(t_k)=c_s(t_s)\left[1-\frac{m^2c_s(t_s)}{3H^2}\ln\left(\frac{k}{k_s}\right)\right]^{-1}~.
\ee
Further, by taking $p\sin\theta\simeq1$, we obtain
\be \label{specificAk}
|A(k)| \simeq \frac{18.5 \times 10^{-5} M_{pl} }{\chi(t_k)}\left[1-\frac{m^2c_s(t_s)}{3H^2}\ln\left(\frac{k}{k_s}\right)\right]^{3/2} ~.
\ee

For the large scale mode with$\ell<60$, the CMB observation yields $A=0.072\pm0.022$. Considering $k\sim l/x_{ls}\simeq7.5\times10^{-5}\ell$, we take $A=0.072$ at $k_a \sim 0.0045 {\rm Mpc}^{-1}$, which exits the Hubble radius at $t_{k_a}$. Therefore, we obtain
\be \label{schi}
\chi(t_{k_a})\simeq 4.38 \times 10^{-3} ~ p M_{pl}\sin\theta ~.
\ee
However, for the small scale mode with $k_b \sim 1 {\rm Mpc}^{-1}$, which exits the Hubble radius at $t_{k_b}$, the quasar observation indicates $A\sim0.0153 $. Correspondingly, we derive
\be \label{lchi}
\chi(t_{k_b}) \simeq 4.25 \times 10^{-3} ~ p M_{pl}\sin\theta ~.
\ee
Since the large scale mode exits the Hubble radius much earlier than the small scale mode, the calculation above suggests that during inflation the $\chi$ field evolves from large values to small ones, which happens to be the picture of large field inflation.

\begin{figure}[tbhp]
\centering
\includegraphics[width=0.45\linewidth]{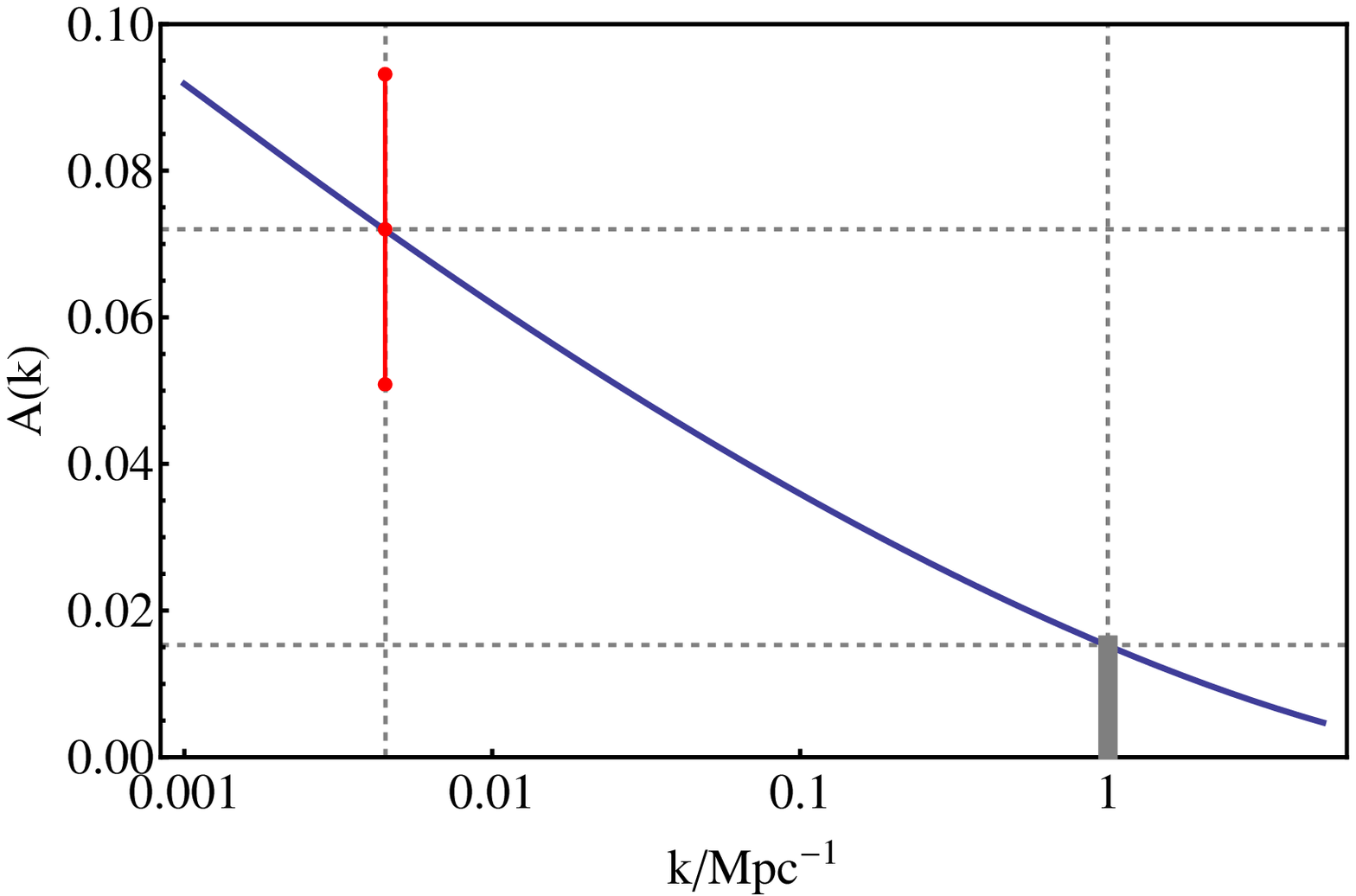}~~~~~~
\includegraphics[width=0.463\linewidth]{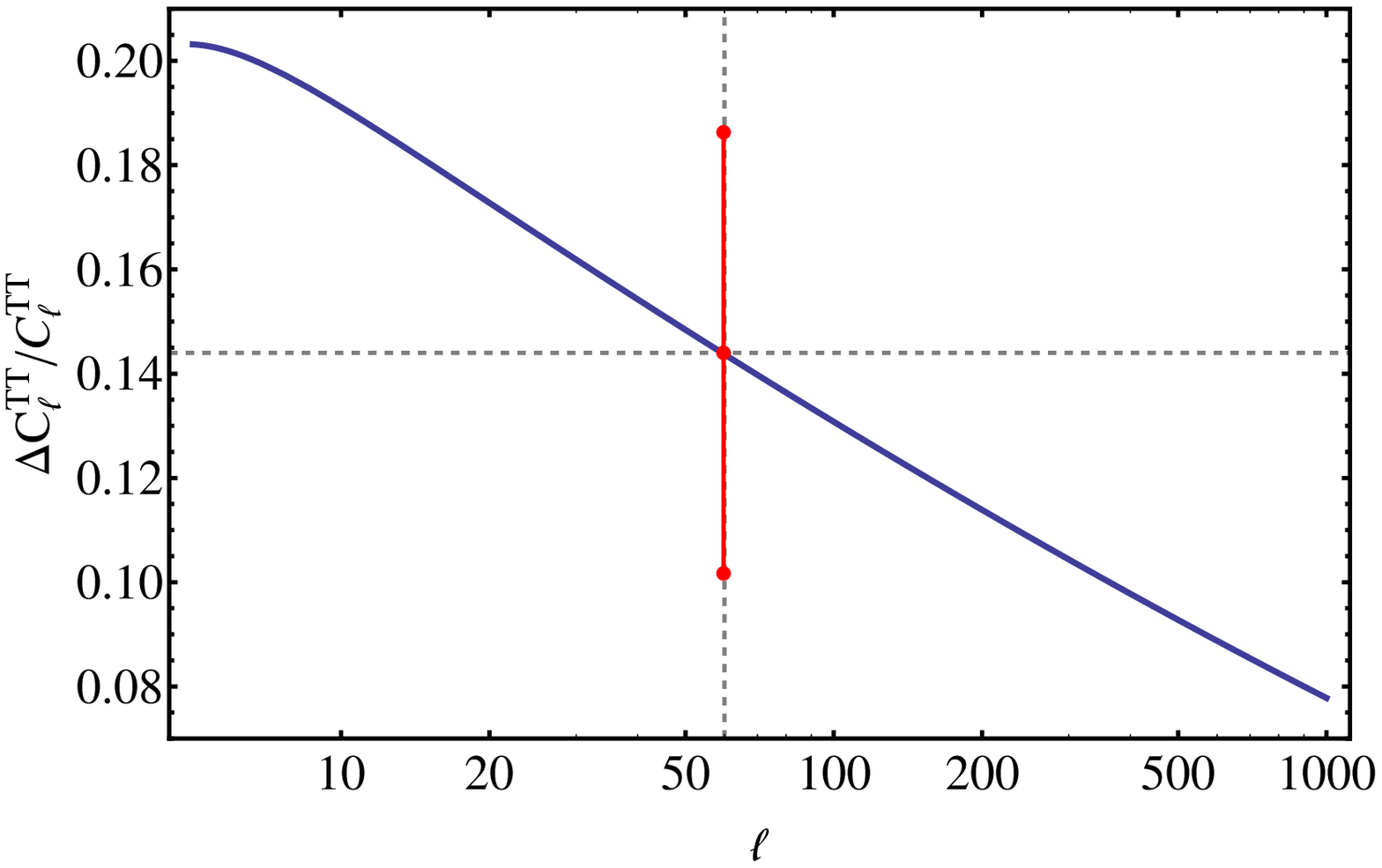}
\caption{Left: The asymmetry amplitude $A$ varies with the comoving wave number $k$. Right: The fractional changes $\Delta C_{\ell}^{\rm TT}/C_{\ell}^{\rm TT}$ in the CMB TT spectrum. The shadow and red error bar correspond to the observational constraints at different scales, which fix the undetermined coefficients. }
\label{Akfig}
\end{figure}

To realize this picture, we turn to study the evolution of $\chi$. Since it is a light field during inflation, we can neglect the effect of potential term, i.e. $V(\chi)\simeq0$. Thus the equation of motion of this canonical field is $\ddot\chi+3H\dot\chi=0$, which yields
\be
\chi(t)=C-De^{-3Ht}~,
\ee
for a nearly constant Hubble constant. Here $C$ and $D$ are the integral constants and can be determined by Eq.(\ref{schi}) and Eq.(\ref{lchi}). With this solution and horizon exit condition, we have $\chi(t_k)=C-D'k^{-3}$. Then the asymmetry amplitude in Eq.(\ref{specificAk}) is determined, whose scale-dependence is shown in the left panel of Fig.\ref{Akfig}. Furthermore, we can also demonstrate the variation in CMB temperature power spectrum by $C^{TT}_{\ell}=\int\frac{dk}{k}P_\zeta\Delta_{T\ell}^2(k)$, where $\Delta_{T\ell}(k)$ is transfer function. Then
\be
\frac{\Delta C^{TT}_{\ell}}{C^{TT}_{\ell}}=\frac{\int2A(k)P_\zeta\Delta_{T\ell}^2(k){dk}/{k}}{\int P_\zeta\Delta_{T\ell}^2(k){dk}/{k}}~.
\ee
Considering that the transfer function is sharply peaked at $\ell=kx_{ls}$, we plot the variation of the TT spectrum in the right panel of Fig.\ref{Akfig}.

\section{Conclusion}

The hemispherical asymmetry of the CMB temperature fluctuations has drawn a lot of attention of cosmologists. This anomaly, if confirmed, will be an important result which is believed to connect with some unknown new physics. Thus it is interesting to explore plausible theoretical explanations. In Ref.~\cite{Cai:2013gma}, it was found that a statistically inhomogeneous sound speed parameter can lead to the spatial variation of the primordial spectrum and a rough calculation of the asymmetry amplitude was provided.

In the present paper, we firstly review the model of multi-speed inflation and the basic ideas of $\delta N$ formalism. Then, we generalize the $\delta N$ formalism to include consistently the contribution of the spatially varying sound speed. The modulation of $c_s$ can be caused by a perturbation mode of very long wavelength seeded by a light field which is not necessary to contribute to the background evolution as well as the generation of curvature perturbations during inflation. And then in the framework of the multi-speed inflation, we give a systematic calculation of the asymmetry amplitude. We find that our approach is free from the constraints of the local non-Gaussianity. For some particular choices of model parameters, we can deduce back to the result derived in \cite{Cai:2013gma}. However, for a more generic case, the behavior of the spatial variation presents significantly differences from the previous result. We point out that our solution may provide a plausible explanation of the scale dependence of the asymmetry amplitude, which accommodates the observations of both CMB and quasar.

Finally we would like to remark that the generalized $\delta N$ formalism, which has taken into account the modulation of primordial sound speed parameter, can be applied to any early universe models involving nonstandard kinetic terms. Therefore, besides the specific multi-speed model we considered in the present paper, it is also interesting to further study inflation models with other k-essence fields.

\acknowledgments
We are grateful to Francis Duplessis, Chunshan Lin, Misao Sasaki, and Yi Wang for valuable comments.
YFC is supported in part by the Chinese National Youth Thousand Talents Program and by the USTC start-up funding (Grant No.~KY2030000049).
DGW is supported in part by the USTC program of national science talent training base in astrophysics.
WZ is supported in part by Project 973 (Grant No. 2012CB821804), by the National Natural Science Foundation of China (Grant Nos. 11173021, 11322324, 11421303), and by project of KIP and CAS.
YZ is supported in part by the National Natural Science Foundation of China (Grant Nos. 11275187, 11421303), by SRFDP, and by the Strategic Priority Research Program ``The Emergence of Cosmological Structures'' of the Chinese Academy of Sciences (Grant No. XDB09000000).

\end{document}